\documentclass[aps,prb,twocolumn]{revtex4}
\usepackage{amsmath}
\usepackage{graphicx}
\usepackage{latexsym}

\DeclareMathOperator{\im}{Im}

\def \j{\vec j}
\def \q{\vec q}

\def \v{\vec v}

\def \be{\begin{equation}}
\def \ee{\end{equation}}
\def \ba{\begin{align*}}
\def \ea{\end{align*}}
\def \ben{\begin{eqnarray}}
\def \een{\end{eqnarray}}

\begin{document}
\title{Investigating superconductor-insulator transition in thin
films using drag resistance:\\
Theoretical analysis of a proposed experiment}
\author{Yue Zou${}^1$, Gil Refael${}^1$, and Jongsoo Yoon${}^2$}
\affiliation{${}^1$Department of Physics, California Institute of
Technology, Pasadena, California 91125, USA \\ ${}^2$Department of Physics, University of Virginia, Charlottesville, Virginia 22903, USA}
\date{\today}

\begin{abstract}

The magnetically driven superconductor-insulator transition in
amorphous thin films (e.g., InO, Ta) exhibits several mysterious phenomena,
such as a putative metallic phase and a huge magnetoresistance
peak. Unfortunately, several conflicting categories of theories, particularly
quantum-vortex condensation, and normal region
percolation, explain key observations equally well. We propose a new
experimental setup, an amorphous thin-film bilayer, where a drag
resistance measurement would clarify the role quantum vortices play
in the transition, and hence decisively point to the correct
picture. We provide a thorough analysis of the device, which
shows that the vortex paradigm gives rise to a drag with an opposite
sign and orders of magnitude larger than the drag measured if
competing paradigms apply.

\end{abstract}
\maketitle

\begin{figure}[h]
\centering
\includegraphics[scale=0.35]{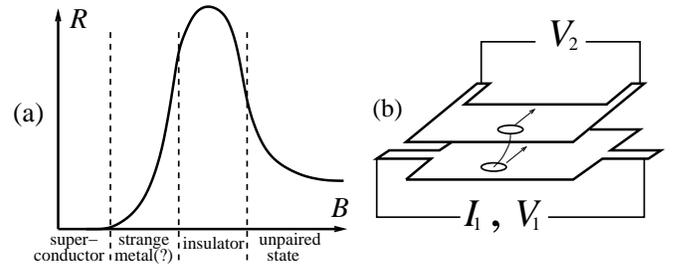}

\caption{(a) A typical magnetoresistance (MR) curve of amorphous thin film superconductors. As the magnetic field $B$ increases, the superconducting phase is destroyed, and a possible metallic phase emerges. After which the system enters an insulating phase, where the MR reaches its peak. The resistance drops down and approaches normal state value as $B$ is further increased. (b) Our proposed bilayer setup for the drag resistance measurement. A current bias $I_1$ is applied in one layer, and a voltage $V_2$ is measured in the other layer. The drag resistance $R_D$ is defined as $R_D=V_2/I_1$.}\label{fig1}
\end{figure}

The superconducting state and the metallic
Fermi-liquid form the very basis of our understanding of correlated
electron systems. Nevertheless, the transition between these two
phases in disordered films is shrouded in mystery.
Experiments probing this transition in amorphous thin films such as Ta,
MoGe, InO, TiN, etc., used a perpendicular magnetic field and
disorder (tuned through film thickness) to destroy
superconductivity. But instead of a superconductor-metal transition,
they observed in many cases a superconductor-insulator-transition (SIT)\cite{earlySIT}.
The "dirty boson" model\cite{Fisher1990} propounded the notion that
the insulator is the mark of vortex condensation, and that the SIT
occurs at a universal critical resistance,
$R_{\Box}=h/4e^2$. More recent experiments, however, showed the
critical resistance to be non-universal
\cite{Kapitulnik2007}. Furthermore, in many field tuned experiments, a surprising metallic phase
intervenes between the superconductor and insulator \cite{Kapitulnik1996,shahar2004,Steiner2005},
with a temperature-independent resistance below $T\sim 50mK$, and (at
least in Ta films) a distinct nonlinear $I-V$ characteristics
\cite{Yoon2005}. Quite generically\cite{shahar2004,Baturina2004,Steiner2005}, these films exhibit a peak in the
magnetoresistance(MR) curve (particularly strong in InO and TiN) as in FIG. \ref{fig1}a.

Two competing categories of theories may account for these phenomena. On one
hand, within the quantum vortex pictures
\cite{Fisher1990,Feigelman1993,vortexmetal}, the insulating phase implies vortex condensation,
the intervening metallic phase is described as uncondensed vortex liquid (e.g., vortex Fermi liquid), and the high field nonmonotonic
MR indicates the appearance of a finite electronic density of
states at the Fermi level. On the other hand, the
percolation  paradigm \cite{Shimshoni1998,meir} describes the films
as consisting of superconducting (SC) and normal puddles; at the MR
peak SC puddles exhibit a Coulomb blockade, and the percolating normal regions
consist of narrow conduction channels.  Yet a third theory tries to account for the low field
SC-metal transition using a phase glass model
\cite{phaseGlass} (see, however, Ref. \onlinecite{Ikeda2007} which argues against these
results), but does not address the full
MR curve. Qualitatively, both
paradigms above are consistent with MR observations, and
recent tilted field, AC conductance, Nernst
effect, and Scanning Tunneling
Spectroscopic measurements\cite{newExperiments} cannot
distinguish between them.

\begin{figure}[h]
\includegraphics[scale=0.36]{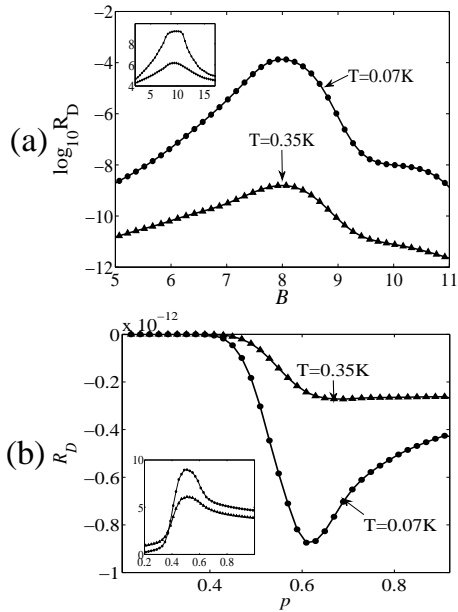}
\caption{Drag resistance $R_D$ (in Ohms) between two
identical films as in FIG.~2b of Ref. \onlinecite{shahar2004} (a)
vs. magnetic field $B$, according to the vortex
picture\cite{vortexmetal} (log scale); (b) vs. normal metal
percentage $p$ (corresponding to magnetic field), according to the
percolation picture\cite{meir}. The drag resistance in (a) has been
smoothened to avoid discontinuity at the boundary between the
metallic and the insulating phase. Center-to-center layer separation
$a=25$nm, temperature $T=0.07$K and 0.35K. Insets: single layer
magnetoresistance (MR, log scale) reproduced in each theory. The
parameters are tuned to make the MR resemble the experimental data
in FIG. 2b of Ref. \onlinecite{shahar2004}.
In the quantum vortex picture, $R_D$ has a peak at the steepest point
($\sim8$T) of the MR, which is due to the fact that $R_D$ is
proportional to the square of the slope of the MR in the small
magnetic field side of the peak. Also, $R_D$ is larger at lower
temperature, because the MR curve is then much steeper. Carrying out the experiments at even lower
temperatures may further enhance the vortex drag effect. In the percolation
picture, the sign of the voltage drop of the passive layer is opposite
to that of the driving layer, and the maximum magnitude value of $R_D$ is
much smaller, $\sim10^{-12}\Omega$.}\label{drag}

\end{figure}

Can we design an experiment that qualitatively distinguishes between
the two paradigms? Here we propose a thin film "Giaever transformer"\cite{Giaever1}
as such an experiment (FIG. \ref{fig1}(b)). The original design of a
Giaever transformer consists of two type-II superconductors
separated by an insulating layer in perpendicular magnetic fields. A current in one layer moves the
vortex lattice in the entire junction, yielding the same DC voltage in
both layers. Determining the drag resistance $R_D=V_2/I_1$ in a similar bilayer structure of two amorphous
superconducting thin films should qualitatively distinguish between the two
paradigms (see also Refs. \cite{Michaeli2006}): Within the vortex paradigm,
vortices in one layer drag
the vortices in the other, but within the percolation picture, the
drag resistance is solely due to
interlayer "Coulomb drag", as studied in semiconductor
heterostructures \cite{Gramila}. The sign and the magnitude of the drag within
 the two paradigms are different: vortex drag implies the same sign for
 the voltage drops in the two layers, $sign(V_1)=sign(V_2)$, but the
 Coulomb drag yields an opposite sign for $V_2$ and $V_1$. In
 addition, a vortex drag would be much stronger than a Coulomb drag,
 because the films' charge carrier density is orders of
magnitude larger than the vortex density; drag effects are typically
inversely correlated with carrier density. Indeed, we find that two
identical films as in FIG. 2b of Ref. \onlinecite{shahar2004} with $25$nm
center-to-center layer separation at $0.07$K would produce a drag resistance
$\sim$ 0.1m$\Omega$ according to the vortex theory, but only
$\sim 10^{-12}\Omega$ for the percolation theory (FIG.\ref{drag}).
Below we will support these claims by
analyzing the drag in the thin film symmetric bilayer within a representative
theoretical framework in the vortex \cite{vortexmetal} and percolation
paradigms \cite{meir}.

{\it Drag resistance, $R_D$, in the quantum vortex picture.}
Within the quantum vortex paradigm, the insulating phase has been explained as a superfluid of vortices by the "dirty boson" model of Ref. \onlinecite{Fisher1990}, while the metallic phase is expected to be an uncondensed vortex liquid. This picture has been pursued by Ref. \onlinecite{vortexmetal} which argues that vortices form a Fermi liquid for a range
of magnetic field, thereby explaining the metallic phase.
At fields larger than the insulating phase value, it is claimed that spinons
(unpaired fermions with finite Fermi-energy DOS) become mobile, impede
vortex motions, destroy the insulating phase, and suppress the resistance down to
normal metallic values.

Without interlayer Josephson coupling, the vortex drag comes from the magnetic coupling between vortices in different layers which tends to align themselves vertically
to minimize the magnetic energy.
To calculate the vortex drag, we follow the vortex-boson duality formalism of Ref. \onlinecite{Fisher1990} and include the effect of physical electromagnetic field to obtain the Lagrangian for vortices,
which features the following form of the intralayer vortex interaction potential $U_i$ and the interlayer one $U_e$:
\be
\begin{aligned}\label{U}
U_i(q)
&=\frac{\phi_0^2q_c}{2\pi}\frac{q+q_c}{q(q^2+2q_cq+q_c^2(1-e^{-2qa}))}\\
U_e(q)
&=-e^{-qa}U_i(q)q_c/(q+q_c),
\end{aligned}
\ee
where $q_c={d}/{(2\lambda^2)}$
is the inverse Pearl penetration depth, $d$ is film thickness, and
$a$ is the center-to-center interlayer distance (we verified that
accounting for the finite thickness gives roughly the same results as
simply taking the interlayer distance to be a center-to-center
one).   $q_c$ can be determined from the
Kosterlitz-Thouless temperature, $T_{KT}$, of the sample; typically
$q_c^{-1}\sim1$cm. When $r<1/q_c$, $U_i(r)$ gives the
familiar log interaction; for $r>1/q_c$, $U_i(r)$ is still
logarithmic but with half of the magnitude \cite{Blatter2005}, in
contrast to the $1/r$ behavior of the single layer case (which is Eq. (\ref{U}) with $a\rightarrow\infty$). The interaction between two vortices
with the same vorticity in different layers is attractive as expected, although
its strength is suppressed with increasing distance $a$ and decreasing $q_c$. These forms of vortex interaction potentials, which can be also derived classically by solving London equations for two vortices, are simply due to the magnetic energy and the superfluid kinetic energy of a SC thin film, and thus is robust against model details.

Deep in the insulating phase, i.e., the vortex superfluid phase where the vortex dynamics is presumably nondissipative, we find the drag resistance using the bilayer supercurrent drag mechanism
\cite{duan1993}, applied to the vortex condensate. Here, a vortex
"supercurrent" $j_1$ in the first layer, produces a vortex
"supercurrent" $j_2$ in the second layer even without
interlayer tunneling. To see this, we use the dual vortex theory
of Ref. \onlinecite{Fisher1990} applied to a bilayer, and
neglect dual gauge field fluctuations, which are suppressed in the
insulating phase. Without vortex current bias in either layer, we can
write a Hamiltonian describing a vortex-superfluid in each layer, with
density-density interaction given by Eqs. (\ref{U}), and use the Bogoliubov approach to diagonalize the Hamiltonian. With a perturbation term
$H_1=\sum_{\q}m_v\j_1\cdot\vec{v_1}$, describing
a vortex current bias with velocity $\v_1$ in the first layer, we can perturbatively find the new ground state and therefore
the drag vortex current ${j_2}$ in the second layer. From
the ratio of the vortex currents in the two layers, which gives the
ratio between the voltages in the two layers,
we find the drag resistance in terms of the single
layer resistance. A straightforward, but lengthy calculation yields:
\be\label{bosonic_drag}
\begin{aligned}
\frac{R_D}R&=\frac{j_2}{j_1}=\frac{\hbar}{128a^2\phi_0}\sqrt{\frac{q_c^3}{2\pi m_vn_v^3}},
\end{aligned}
\ee
where $n_v=B/\phi_0,\,m_v$ are the vortex density and
mass. We expect that in this phase the dissipative
response of vortices, if present, is irrelevant or insufficient to localize
vortices, e.g., as in the Caldeira-Leggett model of a dissipative Josephson junction\cite{SchonZaikin1990}.

In the vortex picture, the intervening metallic phase is interpreted as a liquid of uncondensed vortices, e.g., in Ref. \onlinecite{vortexmetal}, and the vortices have dissipative dynamics. As long as vortex superfluid is absent and the vortex response is dissipative,
one can derive the following form of the the drag resistance $R_D$ using either the Boltzman equation or diagrammatic techniques, irrespective of the effective statistics of vortices\cite{Jauho1993,Oppen2001}:
\be\label{metallic_drag}
R_D=\frac{e^2\phi_0^2}{8\pi^4T}\frac{\partial R_1}{\partial B}\frac{\partial R_2}{\partial B}\int_0^{\infty}q^3dq\int_0^{\infty}
d\omega|U|^2\frac{\im\chi_1\im\chi_2}{\sinh^2\left(\frac{\hbar\omega}{2T}\right)},
\ee
where
$
U={U_e}/\left[{(1+U_{i,1}\chi_1)(1+U_{i,2}\chi_2)-U_e^2\chi_1\chi_2}\right]
$
is the screened interlayer interaction, $\chi_{1,2}$ are the vortex
density response function of each layer. Remarkably, the drag resistance is
proportional to $\partial R_{1,2}/\partial B$, which equals $\partial\sigma_v/\partial n_v$ with $\sigma_v$ and $n_v$ being
respectively the vortices' {\it conductance} and
density. Thus $R_D$ peaks when the MR attains its biggest slope. $\partial\sigma_v/\partial n_v$ appears since $R_D$ is related to the
single layer rectification function, $\Gamma$, defined as $\vec j_v=\Gamma
\phi^2$, with $\phi$ being the vortex potential field. $\Gamma$ is generally
proportional to $\partial\sigma_v/\partial n_v$ (see
Ref. \onlinecite{Oppen2001}). The only model-dependent input is the
density response functions $\chi_{1,2}$. As one choice of
$\chi_{1,2}$, we follow the vortex Fermi liquid description for the
metallic phase\cite{vortexmetal} and use the Hubbard approximation form
for $\chi_{1,2}$ considering the short-range repulsion between
vortices and also the low density of this vortex Fermi
liquid\cite{Hwang2003}. The maximum of $R_D$ we obtained is
$\sim$0.1m$\Omega$ (FIG. \ref{drag}). Note that this result does not
crucially depend on the effective statistics of vortices. We have also
computed $R_D$ by modeling the metallic phase as a classical hard-disk
liquid of vortices\cite{Leutheusser1983}, and the resulting magnitude
and the behavior of $R_D$ is extremely close to the results we
obtained within the vortex Fermi liquid framework\cite{note1}.

At fields larger than the insulating phase or the MR peak, spinons (unpaired electrons) delocalize and impede
vortex motion and suppress drag resistance. Following Ref. \onlinecite{vortexmetal}, we use a semiclassical Drude formalism with statistical interactions between vortices, spinons, and Cooper pairs built in (e.g., the Magnus force on vortices when Cooper pairs move, etc.), and we find that with a
finite spinon conductance, $R_{s}^{-1}>0$,
\be
R_{D}={R_D^0}/\left[{(1+R_v/R_s)^2}\right],
\ee
where $R_D^0$ is the vortex drag resistance with localized spinons,
and $R_v=(h/2e)^2\sigma_v$ is the vortex contribution to the
resistance. Thus, when $R_s\ll R_v$, the drag resistance is quickly
suppressed to immeasurably small as spinon mobility increases.

Lastly, we must estimate the vortex mass
$m_v$. Since there is still controversy over its theoretical value, we chose to estimate it from the experimentally measured activation gap
in the insulating phase\cite{shahar2004,Steiner2005}. When vortices condense, Cooper
pair density fluctuations become gapped due to Higgs mechanism,
and the gap which can be read off from the Lagrangian depends on $m_v$. We conjecture that this gap
is the activation gap. We find
\be\label{m_v}
m_v={8\pi\hbar^2 n_vT_{KT}}/{E_{gap}^2},
\ee
with $n_v=B/\phi_0$. For the InO film of
Ref. \cite{shahar2004}, $T_{KT}\approx0.5$K, and
$E_{gap}\approx1.6$K at $B=9$T, which implies $m_v\approx19m_e$, $m_e$ being the
electron mass in vacuum.
Note that Ref. \onlinecite{shahar2004,Steiner2005} have reported the
suppression of the ratio $T_{KT}/E_{gap}$ by increasing disorder. This
is natural from (\ref{m_v}), since
$T_{KT}/E_{gap}\sim\sqrt{T_{KT}m_v/n_v}$, and disorder suppresses both
$m_v$ and $T_{KT}$. For comparison, this $m_v$ is close to some theoretical results $m_v\sim m_ek_Fd$ for dirty superconductors\cite{Sonin1998}. All analysis above combines to yield the drag resistance behavior,
which we plot in FIG.\ref{drag}(a) for the film as in FIG. 2b of Ref. \onlinecite{shahar2004}.

{\it Drag resistance within the percolation paradigm.} Within the
percolation picture, the non-monotonic
MR arises from the film breaking down to SC and normal regions (described as localized electron glass)
\cite{meir}. As the magnetic field increases, the SC region shrinks,
and a percolation transition occurs. Once the normal regions
percolate, electrons must try to enter a SC island in
pairs, and therefore encounter a large Coulomb blockade absent in normal puddles. The MR peak thus
reflect the competition between electron transport though narrow normal regions, and the tunneling through
SC islands. This picture is captured using a
resistor network description. Each site of the network has a probability $p$ to be
a normal ($1-p$ to be SC); each link is assigned a
resistance from the three values $R_{NN},\,R_{SS},\,R_{SN}$, that
reflect whether the sites the link connects are
normal (N), or superconducting (S). An increase of the magnetic field is
assumed to only cause $p$ to increase. The important ingredient is
that $R_{SN}$ has an activated form with a large gap representing the
charging energy of the SC puddle; $R_{NN}$ reflects the resistance of a localized electron glass with hopping conductivity, and
$R_{SS}$ is mostly negligible.

To calculate $R_D$, we first tune the parameters to make
the single layer resistance resemble the experimental data in
FIG. 2(b) of Ref. \onlinecite{shahar2004}. Next, we place one such
network (active layer) on top of another one (passive layer). Each
link is treated as a subsystem, which might induce a drag voltage (an
emf) $\varepsilon=I R_D$ in the link under it in the passive
layer. When a link is between two normal (SC) sites, it is
treated as a disorder localized electron glass (superconductor). The small
resistance for the SC islands in this theory implies that
vortices in the SC islands, if any, have very low mobility. We find that these vortices have negligible effect on $R_D$\cite{note2}, and Coulomb interaction provide the major
drag effect (more theoretical details will be published elsewhere). Thus, two vertically aligned normal-normal (NN)
links dominate the drag effect. With the electron counterpart of (\ref{metallic_drag}) and the form of electron density response function from Refs. \onlinecite{Shimshoni1997}, we find $R_D$ between
two localized electron glass separated by vacuum is:
\be\label{PercolationDrag}
R_D\approx\frac1{96\pi^2}\frac{R_1R_2}{\hbar/e^2}\frac{T^2}{(e^2nad)^2}\ln\frac1{2x_0}.
\ee
Here, $n\approx5\times10^{20}$cm$^{-3}$ is the typical carrier density of InO\cite{Steiner2005}, $d=20$nm is the film thickness, $a=25$nm is the center-to-center layer separation, $R_{1,2}$ are the resistances of the two NN links, $x_0=a/(2\pi e^2\nu d\xi_L^2)$ where $\nu$ is the density of states and $\xi_L\approx1$nm is the localization length. In deriving
Eq. (\ref{PercolationDrag}) we used the averaged value of the inter-layer coulomb
interaction along the z-direction of the layers. Solving the Kirchoff's equations for the two
layers, we obtain the voltage drop and thereby the drag resistance. The
results are shown in FIG.\ref{drag}(b).

{\it Drag resistance in the phase glass theory.} The phase glass
model\cite{phaseGlass} focuses on the low field SC-metal
transition. It, therefore, does not allow yet a full calculation of
the drag resistance. We leave a full analysis of drag within this
theory for future work, but simply
observe that in the glassy state the phases are ordered locally, and thus
have no mobile vortices. The current-current coupling effects should
therefore be absent, and the
drag is mainly produced by Coulomb interaction. Thus we expect the
sign of the drag voltage to be negative, and the drag resistance should
be small due to the scarcity of excitations in a bosonic system.

{\it Summary.} We have calculated drag resistance in bilayer amorphous thin films separated by an insulator. Our calculation
was carried out within the two competing paradigms, vortex and
percolation pictures, that may account for the phenomena observed at
the breakdown of the superconductivity in amorphous thin films. In the
percolation pictures, the drag resistance is due to interlayer Coulomb
drag and immeasurably small, $\sim 10^{-12}\Omega$. In the
vortex picture, however, the drag is caused by vortex motion. Since the vortex
density is much lower than the charge carrier density, the drag
resistance is orders of magnitude larger; our calculation shows
that it reaches 0.1m$\Omega$
with the same sign as the single layer resistance. These estimates are
made using parameter values that can easily be realized in
experiments. Thus, the drag resistance measurement, albeit challenging
due to the small scale of the maximum drag, can indeed provide
a sharp distinction between competing theoretical paradigms. In our future work, we will incorporate an interlayer Josephson coupling
and analyze its effect on the drag resistance within the different
paradigms. We expect that the drag resistance in both picture will be
enhanced, but the magnitude difference will remain. This would make
 the drag resistance easier to measure, and may not only improve its chances of
 determining the correct theoretical paradigm, but also serve as a
 complementary tool in the quantitative investigation of these
 fascinating systems.
We note that yet another interesting possibility, which we leave for
future research, is to enhance vortex drag by using high magnetic
permeability insulators between the two layers.

It is a pleasure to thank Yonatan Dubi, Jim Eisenstein, Alexander
Finkel'stein, Alex Kamenev, Yen-Hsiang Lin, Yigal Meir, Yuval Oreg, Philip Phillips, Ady Stern, Jiansheng Wu, and Ke Xu
for stimulating discussions. This work was supported by the Research
Corporation's Cottrell award (G.R.), and by NSF through Grant No. DMR-0239450 (J.Y.).

\bibliography{reference}

\begin{thebibliography}{38}
\bibitem[{not()}]{note1}
\bibinfo{howpublished}{The density response function of a classical hard-disk liquid $\chi(k,\omega)$ \cite{Leutheusser1983} in the hydrodynamic limit is given by $
\chi(k,\omega)/\chi(k)=\left(1-1/{\gamma}\right){Dk^2}/({Dk^2-i\omega})+1/{\gamma}\cdot[{c^2k^2-i\omega Dk^2(\gamma-1)}]/({c^2k^2-\omega^2-i\Gamma k^2\omega})$, where $\chi(k)$ is the static compressibility. All parameterss in this formula (including $\chi(k)$) can be determined by two inputs: the diameter of the hard disk vortex, which we use the coherence length, and the vortex mass, which we use the value discussed in the text.}
\end{thebibliography}

\end{document}